\documentclass[12pt]{article}
\pdfoutput=1
\usepackage{cite}
\usepackage[T1]{fontenc}
\usepackage[utf8]{inputenc}

\usepackage[english]{babel}

\usepackage{graphicx}
\usepackage[dvipsnames]{xcolor}

\usepackage{amsmath}
\usepackage{amsfonts}
\usepackage{amssymb}
\usepackage{amstext}
\usepackage{slashed}

\usepackage[bookmarks, breaklinks, colorlinks,urlcolor=black, citecolor=red, 
linkcolor=blue]{hyperref}

\usepackage{multirow}
\usepackage{booktabs}

\usepackage{hyperref}
\usepackage{placeins}
\usepackage{verbatim}

\usepackage{mathtools}
\usepackage{float} 

\def\dsp{\displaystyle}
\def\bc{\begin{center}}
\def\ec{\end{center}}
\def\be{\begin{equation}}
\def\ee{\end{equation}}
\def\bea{\begin{eqnarray}}
\def\eea{\end{eqnarray}} 
\def\nn {\nonumber}

\def\gev{\ensuremath{\mathrm{Ge\kern -0.1em V}}}
\def \Re{\text{Re}}
\def \Im{\text{Im}}

\def \dtwo{\ensuremath{D_2^*}}
\def\thD{{\ensuremath{\theta_D}}}
\def\thl{{\ensuremath{\theta_\ell}}}
\def\AFB{A_{\text{FB}}}
\def \mdtwo{\ensuremath{m_{D_2^*}}}

\def \azeL{{{\cal A}_0^L}}

\def \apaL{{{\cal A}_\parallel^L}}

\def \apeL{{{\cal A}_\perp^L}}

\def \azeTL{{{\cal A}_{T\,0}^L}}

\def \apaTL{{{\cal A}_{T\,\|}^L}}

\def \apeTL{{{\cal A}_{T\,\perp}^L}}

\def\rd {R(D)}
\def\rdst {R(D^*)}
\def\rdrdst {R(D^{(*)})}

\def\rJpsi{R_{J/\psi}}

\usepackage{relsize}
\def\Babar{{\mbox{\slshape B\kern-0.1em{\smaller A}\kern-0.1em B\kern-0.1em{\smaller A\kern-0.2em R}}}}

\begin{document}
	
	\renewcommand*{\thefootnote}{\fnsymbol{footnote}}
	
	
\mbox{}\hfill{{\small SI-HEP-2019-18}}
\vskip 2mm	

	\begin{center}
		
		{\Large\bf{Angular analysis of $\bar{B}\to D_2^*(\to D \pi)\ell \bar{\nu}$ decay \\ and new physics}} \\[6mm]
		{ Rusa Mandal \footnote{Email: Rusa.Mandal@uni-siegen.de}
		}
		
		{\small\em Theoretische Elementarteilchenphysik, Naturwiss.- techn. Fakult$\ddot{a}$t, \\ Universit$\ddot{a}$t Siegen, 57068 Siegen, Germany}

	\end{center}
	
	
	\begin{abstract}
	
	We derive the four-fold angular distribution for the semileptonic decay $\bar{B}\to D_2^*(\to D \pi)\,\ell \bar{\nu}$ where $\dtwo$(2460) is a tensor meson. We start with the most general beyond the Standard Model (SM) dimension-six effective Hamiltonian which comprises (axial)vector, (pseudo)scalar and tensor operators for both quark and lepton currents, and it also includes the right-handed neutrinos. The decay can be described by 16 transversity amplitudes and it provides a multitude of observables which can be extracted from data. We investigate the observables in the context of the SM and the new physics scenarios which can explain the intriguing discrepancies observed in the $b \to c \tau \bar{\nu}$ data. 

	\end{abstract}

\section{Introduction}
\label{sec:Intro}

In the absence of any clean signal of beyond the Standard Model (SM) particle, the effective theory analysis has become one of the primary directions to pursue. In a situation where the scale of new physics (NP) might be quite higher and thus the direct production remains awaiting at the colliders, higher dimensional operators can capture its effect. It is a historical fact that several discoveries in particle physics were preceded by indirect evidence through quantum loop contributions. The decays of $B$ mesons are the important probes for such searches. While the fully hadronic decay modes are subject to large and, in cases, not-so-well understood
strong interaction corrections, the situation is much more under control for semileptonic decays. 

Among such, the $b\to c\ell \bar{\nu}$ modes are of special interests. In the SM, this decay proceeds through a tree level $W$ boson exchange and thus is not suppressed as compared to flavor-changing neutral current transitions. In spite of being a charged current channel, some intriguing hints of discrepancies have been observed by several experimental collaborations.
The ratios of branching fraction (BR) are particularly clean probes of
physics beyond the SM, due to the cancellation of the leading
uncertainties inherent in individual BR predictions, defined as
\bea
\label{eq:RD}
\rdrdst \equiv \frac{ {\rm BR}(B\to D^{(*)}\tau\nu)}{ {\rm BR}(B\to D^{(*)}\ell \nu)}\,,
\eea
with $\ell=e$ or $\mu$, and the ratio $R_{J/\psi}$ defined as
\bea
\label{eq:defrjpsi}
\rJpsi \equiv \frac{ {\rm BR}(B_c\to J/\psi\, \tau\nu)}{ {\rm BR}(B_c\to J/\psi\, \mu \nu)}\,.
\eea
Stating from the first measurement of $\rdrdst$ by \Babar~\cite{Lees:2012xj,Lees:2013uzd} Collaboration and then with subsequent results from Belle~\cite{Huschle:2015rga,Hirose:2016wfn,Hirose:2017dxl,Abdesselam:2019dgh} and LHCb~\cite{Aaij:2015yra,Aaij:2017uff} in recent years, the latest averages performed by the HFLAV Group~\cite{hflav} point $1.4\sigma$ and $2.5\sigma$ deviations for $\rd$ and $\rdst$, respectively. In each case, the data exceeds the SM estimate and while considering the $\rd-\rdst$ correlation, the resulting discrepancy at a $3\sigma$ level.
In the case for the ratio $\rJpsi$, which has the same quark-level transition $b \to c \ell \bar{\nu}$, in the $B_c$ meson system, the LHCb Collaboration has observed the hint of another discrepancy~\cite{Aaij:2017tyk} within $2\sigma$ level. While these hints are subject to confirmation or falsification with more statistics, efforts to explore additional modes mediated by the same parton-level transition should be open ended.

In this paper we study the decay having the same $b \to c \ell \bar{\nu}$ transition but with a tensor meson in the final state, namely $\bar{B} \to \dtwo \ell \bar{\nu}$. The $\dtwo(2460)$ meson has $J^P=2^+$ and it dominantly decays to $D\pi$ final states. The currently available data for $\dtwo$ is from Belle~\cite{Liventsev:2007rb} and \Babar~\cite{Aubert:2008zc} Collaborations on the product of branching  fractions as ${\rm BR}(B\rightarrow {D}_2^*\ell^{}\overline{\nu}_{\ell})\times {\rm BR} ({D}_2^*\rightarrow D\pi)\simeq \mathcal{O}(10^{-3})$. Thus the analysis of the semileptonic decay $\bar{B} \to \dtwo(\to D \pi) \ell \bar{\nu}$ not only provides a complementary information about the possible NP evidence, but also constitutes backgrounds to the $\rdrdst$ measurements. The decay rate for this channel has been calculated using the heavy quark effective theory expansion of the form factors including terms in all order in $\Lambda_{QCD}/m_{b,c}$ and $\alpha_s$ in Ref.~\cite{Bernlochner:2017jxt}, with three-point QCD sum rule form factors in Ref.~\cite{Azizi:2013aua} and recently using light-cone sum rules (LCSR) form factors in Ref.~\cite{Aliev:2019ojc}. The ratio of branching fractions for different lepton flavors has also been estimated. In this paper we derive the full four-fold differential distribution in terms of four kinematic variables i.e., three angles and the dilepton invariant mass square, for the most general dimension-six beyond the SM effective Hamiltonian including the right-handed neutrinos. The full angular distribution provides additional information about a the orientations of the final states and the polarizations of the $\dtwo$ meson through multitude of observables. A very well-known example is the $\bar{B}\to K^*(\to K \pi)\ell \bar{\ell}$ decay~\cite{Kruger:1999xa} where all the angular observables are measured at the LHCb with 3\,fb$^{-1}$ data~\cite{Aaij:2015oid}.  In literature several analyses can be found for the decay $\bar{B} \to D^* \ell \bar{\nu}$~\cite{Biancofiore:2013ki,Colangelo:2016ymy,Colangelo:2018cnj,Becirevic:2019tpx,Duraisamy:2013kcw} and $B$ to light mesons~\cite{Colangelo:2019axi} as well. The effect of spin-2 operators in modes with tensor meson can be found in Ref.~\cite{Gratrex:2015hna}

The rest of the paper is organized as following. In Sec.~\ref{sec:Lag} we start with the general effective Hamiltonian and discuss the theoretical framework of the decay. The full angular distribution is derived in Sec.~\ref{sec:dist} and several observables are constructed. We perform the phenomenological analysis of the angular observables in Sec.~\ref{sec:pheno} where variations due to possible presence of NP operators are highlighted. Finally Sec.~\ref{sec:summary} summarizes the results with discussions.

\section{Theoretical framework}
\label{sec:Lag}

The effective theory description of semileptonic $B$ decays requires separation of short-distance (QCD, weak interaction and NP) and long-distance QCD in an effective Hamiltonian. Being a tree-level process, in the SM, the $b\to c \ell \bar{\nu}$ effective Hamiltonian has a very simple form with only one left-handed vector current four-fermion operator. However, physics beyond the SM can contribute via operators with same and/or different Lorentz structures. Thus we start with the most general dimension-six beyond the SM effective Hamiltonian for $b\to c \ell \bar{\nu}$ transition
\begin{align}
\label{eq:Lag}
{\mathcal{H}}_{\rm eff} =&  \frac{4 G_F V_{cb}}{\sqrt{2}}\Big\{ \mathcal{O}^V_{LL} + \sum_{\substack{X=S,V,T \\ M,N=L,R }} C^X_{MN} \mathcal{O}^X_{MN}  \Big\}, \
\end{align}
where the four-fermion operators are defined for $M,N\in \{L,R\}$ as
\begin{align}
 \mathcal{O}_{M N}^{S} & \equiv\left(\bar{c} P_{M} b\right)\left(\bar{\ell} P_{N} \nu\right), \\ \mathcal{O}_{M N}^{V} & \equiv\left(\bar{c} \gamma^{\mu} P_{M} b\right)\left(\bar{\ell}
 \gamma_{\mu} P_{N} \nu\right), \\ 
\mathcal{O}_{M N}^{T} & \equiv\left(\bar{c} \sigma^{\mu \nu} P_{M} b\right)\left(\bar{\ell} \sigma_{\mu \nu} P_{N} \nu\right).
\end{align}
The Wilson coefficients $C^X_{MN}=0$ in the SM and they encode the short-distance physics which can be generated by heavy NP mediators. The neutrino oscillation experiments confirm the tiny mass of neutrinos which in turn implies the neutrinos may not be purely left-handed and for generality we also include the light right-handed neutrinos in Eq.~\eqref{eq:Lag}. It can be shown using Fierz rearrangements, that for $M\neq N$, the tensor operators vanish identically.

As a next step, we need to parametrize the $\bar{B}\to \dtwo$ hadronic matix elements. For a spin-2 particle $\dtwo$, the polarization tensor satisfies $\epsilon_{\mu\nu}
p^{\nu}_{\dtwo}=0$ (with $p^{\nu}_{\dtwo}$ the four-momentum) and is symmetric and
traceless. The polarization tensor $\epsilon_{\mu\nu}$ can be constructed via the spin-1 polarization vector $\epsilon_\mu$:
\begin{eqnarray}
\label{eq:eps1}
&&\epsilon_{\mu\nu}(\pm2)=
\epsilon_\mu(\pm)\epsilon_\nu(\pm),\;\;\;\;
\epsilon_{\mu\nu}(\pm1)=\frac{1}{\sqrt2}
[\epsilon_{\mu}(\pm)\epsilon_\nu(0)+\epsilon_{\nu}(\pm)\epsilon_\mu(0)],\nonumber\\
&&\epsilon_{\mu\nu}(0)=\frac{1}{\sqrt6}
[\epsilon_{\mu}(+)\epsilon_\nu(-)+\epsilon_{\nu}(+)\epsilon_\mu(-)]
+\sqrt{\frac{2}{3}}\epsilon_{\mu}(0)\epsilon_\nu(0).
\end{eqnarray}
In the case of the tensor meson moving along the $z$ axis, the $\epsilon_\mu$ are usually chosen as
\begin{eqnarray}
\label{eq:eps2}
\epsilon_\mu(0)&=&\frac{1}{\mdtwo}(|\vec
p_{D_2^*}|,0,0,E_{D_2^*}),\;\;\;
\epsilon_\mu(\pm)=\frac{1}{\sqrt{2}}(0,\mp1,-i,0),
\end{eqnarray}
where $E_{D_2^*}$ and $\vec{p}_{D_2^*}$ is the energy and the three-momentum of $\dtwo$ in the $B$ meson rest frame, respectively.

The relevant form factors for the $\bar{B} \to \dtwo$ matrix elements of the vector and  axial-vector currents are defined as \cite{Aliev:2019ojc}
\bea
\label{eq:matVA}
\langle \dtwo(p_{\dtwo},\epsilon^*)|\bar{c}\gamma_{\mu}b|  \bar{B}(p_B)\rangle  &=&
\frac{2 i V(q^2)}{m_B + m_{\dtwo}}\varepsilon_{\mu \nu \rho \sigma} \epsilon_T^{*\nu}  p^{\rho}_{\dtwo} p^{\sigma}_B \,,\\
\langle{ \dtwo(p_{\dtwo},\epsilon^*)}| \bar{c}\gamma_{\mu} \gamma_5 b|{\bar{B}(p_B)} \rangle &=&  2 m_{\dtwo} A_0 (q^2)\frac{\epsilon_T^*.q}{q^2} q_\mu + (m_B + m_{\dtwo}) A_1(q^2) \Big[{\epsilon_T^*}_{\mu}-\frac{\epsilon_T^*.q}{q^2} q_\mu \Big]\nn \\ && -A_2(q^2) \frac{\epsilon_T^*.q}{(m_B + m_{\dtwo})} \Big[(p_B +p_{\dtwo})_\mu -\frac{m^2_B-m^2_{\dtwo}}{q^2}q_\mu \Big]\,.
\eea
where $\epsilon_{T}^\mu(h) = \epsilon^{\mu\nu}(h)q_\nu/ m_B$ with $q_\mu=(p_B-p_{\dtwo})_\mu$ being the momentum transfer.

It can be shown that the $\bar{B} \to \dtwo$ matrix element for the scalar current vanishes and the pseudoscalar current reduces to
\bea
\label{eq:matP}
\langle \dtwo(p_{\dtwo},\epsilon^*)|\bar{c}\gamma_5 b| \bar{B}(p_B)\rangle  &=& -\frac{ 2 m_{\dtwo} A_0 (q^2)}{m_b(\mu) + m_c(\mu)}\epsilon^*_T.q\,.
\eea

Next the tensor operators are parametrized with the well-known form factors $T_i$ defined as
\begin{align}
\label{eq:matT}
\langle{ \dtwo(p_{\dtwo},\epsilon^*)}|\bar{c} \sigma_{\mu\nu} q^\nu b|{\bar{B}(p_B)} \rangle =&~ \epsilon_{\mu\nu\rho\sigma}\,\epsilon_T^{*\nu} p_{\dtwo}^\rho\, p_B^\sigma\,2\,T_1(q^2) \,, \\ \nn
\langle \dtwo(p_{\dtwo},\epsilon)|\bar{c} \sigma_{\mu\nu}\gamma_5 q^\nu b|\bar{B}(p_B)\rangle =& \left[(m_B^2-\mdtwo^2)\epsilon_T^{*\mu} - (\epsilon_T^*. q)(p_B +p_{\dtwo})_\mu\right]T_2(q^2) \\
& - (\epsilon_T^* .q)\left[q_\mu-{q^2 \over m_B^2-\mdtwo^2}(p_B +p_{\dtwo})_\mu\right]T_3(q^2) \,.
\end{align}

To express $\bar{B}\to D \pi$ matrix elements in terms of $\bar{B}\to \dtwo$ form factors we assume the $\dtwo$ decays resonantly. This allows us to use the narrow-width approximation for the $\dtwo$ propagator as follows:
$$ \frac{1}{(p_{\dtwo}^2-\mdtwo^2)+ (\mdtwo \Gamma_{\dtwo})^2} \xrightarrow[]{\Gamma_{\dtwo}\ll \mdtwo} \frac{\pi}{\mdtwo \Gamma_{\dtwo}}\, \delta(p_{\dtwo}^2-\mdtwo^2)\,.$$
We can write the hadronic matrix elements in Eqs.~\eqref{eq:matVA}--\eqref{eq:matT} as
\bea
\langle \dtwo(p_{\dtwo},\epsilon^*)|\mathcal{J}^\mu| \bar{B}(p_B)\rangle  &=& \epsilon_{\alpha\beta}^* \, q^\beta A^{\mu \alpha}
\eea
where $A^{\mu \alpha}$ contains the $\bar{B}\to \dtwo$ form factors.
With the effective Lagrangian describing the $\dtwo\to D \pi$ decay 
$$\mathcal{L} = g_{\dtwo D \pi}\, \epsilon_{\mu\nu} p_D^\mu p_\pi^\nu\,, $$
we obtain the total decay width $\Gamma_{\dtwo}= \dsp\frac{g_{\dtwo D \pi}^2 \mdtwo^3}{60 \pi} \dsp\frac{\beta^5}{2^5 }$, where $\beta=\dsp\frac{\lambda^{1/2}(\mdtwo^2,m_D^2,m_\pi^2)}{\mdtwo^2}$ with  $\lambda(x,y,z)\equiv x^2+y^2+z^2-2xy-2xz-2yz$ is the K\"allen function.

Using the standard expression for the sum over polarization tensor $\sum \limits_{m=1}^5 \epsilon^{\mu \nu}(m) \epsilon^{\rho \sigma}(m)=B^{\mu \nu, \rho \sigma}$ with
\begin{align}
B^{\mu \nu, \rho \sigma}(p_{\dtwo})=& \frac{1}{2}\left(\eta^{\mu \rho}-\frac{p_{\dtwo}^{\mu} p_{\dtwo}^{\rho}}{\mdtwo^{2}}\right)\left(\eta^{\nu \sigma}-\frac{p_{\dtwo}^{\nu} p_{\dtwo}^{\sigma}}{\mdtwo^{2}}\right)+\frac{1}{2}\left(\eta^{\mu \sigma}-\frac{p_{\dtwo}^{\mu} p_{\dtwo}^{\sigma}}{\mdtwo^{2}}\right)\left(\eta^{\nu \rho}-\frac{p_{\dtwo}^{\nu} p_{\dtwo}^{\rho}}{\mdtwo^{2}}\right) \nn \\ &-\frac{1}{3}\left(\eta^{\mu \nu}-\frac{p_{\dtwo}^{\mu} p_{\dtwo}^{\nu}}{\mdtwo^{2}}\right)\left(\eta^{\rho \sigma}-\frac{p_{\dtwo}^{\rho} p_{\dtwo}^{\sigma}}{\mdtwo^{2}}\right) 
\end{align}
we get the desired matrix element
\bea
\langle D(p_{D})\pi(p_\pi)|\mathcal{J}^\mu| \bar{B}(p_B)\rangle  &=&  D_{\dtwo}(p_{\dtwo}^2) W_\alpha A^{\mu \alpha}\,,
\eea
where 
\begin{eqnarray}
\label{eq:notation}
|D_{\dtwo}(p_{\dtwo}^2)|^2 &=& \frac{2^5 \times 60 \pi^2}{\beta^5 \mdtwo^4} \, \delta(p_{\dtwo}^2-\mdtwo^2)\,,\nn \\
W^\alpha&=& (P^\alpha - \xi\, p_{\dtwo}^\alpha ) (P.q- \xi\, p_{\dtwo}.q) + \frac{1}{3} \beta^2 \mdtwo^2 \big( q^\alpha - \frac{p_{\dtwo}.q}{\mdtwo^2} p_{\dtwo}^\alpha \big),\\ 
~P^\mu&=& p_D^\mu-p_\pi^\mu,~\xi= \frac{m_D^2-m_\pi^2}{\mdtwo^2}\nn.
\end{eqnarray}
With the framework defined above, we are now in a stage to compute the differential distribution for the $\bar{B}\to \dtwo(\to D \pi)\ell^-\bar{\nu}$ decay.

\section{Angular distribution and observables}
\label{sec:dist}

In this section, we derive the full four-body angular distribution of the semileptonic decay $\bar{B}(p_B)\to \dtwo(p_{\dtwo})\,\ell^-(q_2)\,\bar{\nu}(q_1)$, with $\dtwo(p_{\dtwo})\to D(p_D)\pi(p_\pi)$ on the mass shell. This process is completely described by four independent kinematic variables. These kinematic variables are the lepton-pair invariant mass squared $q^2=(q_1+q_2)^2$, and the three angles $\phi$, $\theta_\ell$ and $\theta_D$.  The angles $\theta_\ell$ and $\theta_D$ are defined as follows: assuming that the $\dtwo$ has a momentum along the positive $z$ direction in the $B$ rest frame, $\theta_D$ is the angle between the $D$ and the
$+z$ axis and $\theta_\ell$ is the angle of the $\ell^-$ with the $+z$ axis. The angle $\phi$ is the angle between the decay planes formed by $\ell^-\bar{\nu}$ and $D\pi$. Squaring the matrix element, summing over spins of the final state particles and
using the kinematical identities given in App.~\ref{app:kin} we obtain the differential decay distribution of $\bar{B}\to \dtwo(\to D\pi)\ell^-\bar{\nu}$ as
\begin{align}
\label{eq:dGamma}
 \frac{d^4\Gamma}{dq^2d\cos\theta_D d\cos\theta_l d\phi}=& \frac{15}{128 \pi}  \Big[I_1^c (3\cos^2\theta_{D}-1)^2 + 3 I_1^s  \sin^2(2\theta_{D}) + I_2^c  (3\cos^2\theta_{D}-1)^2 \cos(2\theta_l) \nn \\
 +& 3I_2^s \sin^2(2\theta_{D}) \cos(2\theta_l) + 3 I_3  \sin^2(2\theta_{D}) \sin^2\theta_l
\cos(2\phi) \nn \\
+& 2\sqrt 3 I_4 (3\cos^2\theta_{D}-1)  \sin(2\theta_{D}) \sin(2\theta_l)\cos\phi \nn \\
+&  2\sqrt 3 I_5 (3\cos^2\theta_{D}-1)  \sin(2\theta_{D}) \sin(\theta_l) \cos\phi \nn \\
+& 3 I_6^s  \sin^2(2\theta_{D}) \cos\theta_l + I_6^c (3\cos^2\theta_{D}-1)^2 \cos\theta_l \nn \\
+& 2\sqrt 3 I_7 (3\cos^2\theta_{D}-1)  \sin(2\theta_{D}) \sin(\theta_l) \sin\phi\nn \\
+& 2\sqrt 3 I_8 (3\cos^2\theta_{D}-1)   \sin(2\theta_{D})\sin(2\theta_l)\sin\phi \nn \\
+& 3 I_9  \sin^2(2\theta_{D}) \sin^2\theta_l \sin(2\phi)\Big].
\end{align}
The angular coefficients $I_i$ are functions of tranversity amplitudes given by
\bea
\label{eq:I1c}
I^c_1 &=&  2 \Big(1 + \frac{m^2_l}{q^2} \Big) \Big(|\azeL|^2 +4|\azeTL|^2\Big) - 16\frac{m_l}{\sqrt{q^2}}\Re[\azeL \azeTL^{\!\!\!\!*}\,]  +  \frac{4 m^2_l}{q^2} |A_{tP}^L|^2 + \left( L \to R \right) \,, 
\eea
\bea
I^s_1 &=&  \frac{1}{2} \Big(3 +  \frac{m^2_l}{q^2} \Big) \Big(|\apeL|^2 + |\apaL|^2  \Big)+ 2 \Big(1 +  \frac{3m^2_l}{q^2} \Big) \Big(|\apeTL|^2 + |\apaTL|^2  \Big)\,, \nn \\
&-&  8\frac{m_l}{\sqrt{q^2}}\Re[\apeL \apeTL^{\!\!\!\!*} + \apaL \apaTL^{\!\!\!\!*}\,]+ \left( L \to R \right),  \\
I^c_2 &=& -2 \Big(1 -  \frac{m^2_l}{q^2} \Big)  \Big(|\azeL|^2 - 4\,|\azeTL|^2  + \left( L \to R \right) \Big), \\
I^s_2 &=&  \frac{1}{2} \Big(1 -  \frac{m^2_l}{q^2} \Big)\Big(|\apeL|^2 + |\apaL|^2 - 4\big(|\apeTL|^2 + |\apaTL|^2 \big)  + \left( L \to R \right) \Big)\,, \\
\label{eq:I3}
I_3 &=& \Big(1 -  \frac{m^2_l}{q^2} \Big) \Big(|\apeL|^2 - |\apaL|^2 -4 \big(|\apeTL|^2 - |\apaTL|^2 \big)+ \left( L \to R \right) \Big)\,, \\
\label{eq:I4}
I_4 &=& - \sqrt{2}  \Big(1 -  \frac{m^2_l}{q^2} \Big) \Big(\Re [\azeL^{}\apaL^* -4\, \azeTL^{}\apaTL^{\hspace{-2mm}*} + \left( L \to R \right)] \Big) \,, \\
\label{eq:I5}
I_5 &=& -2 \sqrt{2} \bigg\{\Re[\big( \azeL^{} -2 \frac{m_l}{\sqrt{q^2}} \azeTL \big) \big( \apeL^* -2 \frac{m_l}{\sqrt{q^2}} \apeTL^{\hspace{-3mm}*}~ \big) -\left( L \to R \right) ] \nn \\
&-&\frac{m^2_l}{q^2}  \Re[ {A^{L}}^*_{\!\!\!\!tP} \big( \apaL - 2\frac{\sqrt{q^2}}{m_l} \apaTL\big)+\left( L \to R \right)  ]\bigg\}\, \\
I^c_6 &=&  8  \frac{m^2_l}{q^2} {\rm Re}[{A^{L}}^*_{\!\!\!\!tP} \big( \azeL - 2\frac{\sqrt{q^2}}{m_l} \azeTL\big) + \left( L \to R \right) ] \,, \\
I^s_6 &=& 4  {\rm Re}[\big( \apaL^{} -2 \frac{m_l}{\sqrt{q^2}} \apaTL\big) \big( \apeL^* -2 \frac{m_l}{\sqrt{q^2}} \apeTL^{\hspace{-3mm}*}~ \big) -\left( L \to R \right) ] \,, \\
I_7 &=&  2 \sqrt{2} \bigg\{\Im [\big( \azeL^{} -2 \frac{m_l}{\sqrt{q^2}} \azeTL \big) \big( \apaL^* -2 \frac{m_l}{\sqrt{q^2}} \apaTL^{\hspace{-2mm}*}~ \big) -\left( L \to R \right)] \nn \\
&+& \frac{m^2_l}{q^2} \Im[  {A^{L}}^*_{\!\!\!\!tP} \big( \apeL - 2\frac{\sqrt{q^2}}{m_l} \apeTL\big) + \left( L \to R \right) ]\bigg\}\,, \\
I_8 &=& -\sqrt{2}  \Big(1 -  \frac{m^2_l}{q^2} \Big) {\rm Im}[\azeL\apeL^{*} -4\, \azeTL \,\apeTL^{{\hspace{-2mm}*}}+ \left( L \to R \right) ]\,, \\
\label{eq:I9}
I_9 &=&  2\Big(1 -  \frac{m^2_l}{q^2} \Big) {\rm Im}[\apaL^{}\apeL^* -4 \apaTL^{}\apeTL^{\hspace{-3mm}*}+ \left( L \to R \right)] \,.
\eea

The transversity amplitudes are the projections of the total decay amplitude into the explicit polarization basis. In the SM, the decay $\bar{B}\to \dtwo\ell^-\bar{\nu}$ can be described by a total of four transversity amplitudes. Notice from Eqs.~\eqref{eq:eps1} and \eqref{eq:eps2} that $\epsilon^{\mu\nu}(\pm 2)q_\nu=0$ in the $B$-rest frame, implying only three states of polarization contribute to the considered decay. As a result we have four transversity amplitudes corresponding to one longitudinal ($\mathcal{A}_0$), two transverse ($\mathcal{A}_{\perp,\|}$) direction, and a timelike amplitude ($\mathcal{A}_t$) for the virtual vector boson decaying to lepton-antineutrino pair. However, inclusion of right-handed neutrinos distinguish the left and right chirality of the leptonic current and we get a total of eight amplitudes $\mathcal{A}^{L,R}_{0,\perp,\|,t}$. Now in the presence of the NP operators given in Eq.~\eqref{eq:Lag}, the new (axial)vector contributions can be incorporated in the above-mentioned eight transversity amplitudes modified with new Wilson coefficients; however, the (pseudo)scalar and tensor operators induce eight further (four for each chirality of leptonic current) amplitudes. These are two (pseudo)scalar amplitudes $\mathcal{A}_P^{L,R}$ and six for the tensor operators $\mathcal{A}^{L,R}_{T\,0,T\,\perp,T\,\|}$. Thus with the most general effective Hamiltonian (Eq.~\eqref{eq:Lag}), the four-body decay can be described by a total of 16 tranversity amplitudes. Note that we have suppressed the $q^2$ dependence in the angular coefficients as well as in the transversity amplitudes and will continue to do so for simplicity.

Defining the normalization factor
\bea
\label{NF}
N_F &=& \Big[ \frac{G^2_F \sqrt{\lambda(m_B^2,{\mdtwo}^2,q^2)} |V_{cb}|^2 q^2 }{3\times 2^{7}\pi^3 m^3_B} \Big(1-\frac{m_l^2}{q^2}\Big)^2 ~Br(\dtwo\rightarrow D\pi)\Big]^{1/2} \;
\eea
and by introducing the following notation for the NP Wilson coefficients
\bea
\label{eq:WC-nom}
C^V_{RL} \pm C^V_{LL} \equiv g_{V,A}^L,~C^V_{RR} \pm C^V_{LR} \equiv g_{V,A}^R,~C^S_{RL} \pm C^S_{LL} \equiv g_{S,P}^L,~C^S_{RR} \pm C^S_{LR} \equiv g_{S,P}^R,
\eea
we write the expressions for the  transversity amplitudes arising from (axial)vector operators as
\begin{align}
\label{tran_amp}
{\mathcal{A}}_0^{L}  &= N_F\,\frac{1}{2 \mdtwo \sqrt{q^2}} \frac{\lambda^{1/2}(m_B^2,\mdtwo^2,q^2)}{\sqrt6 m_B\mdtwo} \Big[(m_B^2 - \mdtwo^2 - q^2) (m_B + \mdtwo ) A_1(q^2) \nn \\
& \hspace*{6cm}-\frac{\lambda(m_B^2,\mdtwo^2,q^2)}{m_B +m_{\dtwo}}  A_2(q^2) \Big]\left(1 - g_A^{L}\right)\,, \\
{\mathcal{A}}_0^{R}  &= N_F\,\frac{1}{2 \mdtwo \sqrt{q^2}} \frac{\lambda^{1/2}(m_B^2,\mdtwo^2,q^2)}{\sqrt6 m_B\mdtwo} \Big[(m_B^2 - \mdtwo^2 - q^2) (m_B + \mdtwo ) A_1(q^2) \nn \\
&\hspace*{6cm} -\frac{\lambda(m_B^2,{\mdtwo}^2,q^2)}{m_B +m_{\dtwo}}  A_2(q^2) \Big]\left(- g_A^{R}\right) \,, \\
{\mathcal{A}}_{\|}^{L}  &= N_F\,\frac{\lambda^{1/2}(m_B^2,\mdtwo^2,q^2)}{2 m_B\mdtwo} \,(m_B + \mdtwo)A_1(q^2)\, \left(1 - g_A^{L}\right)  \,, \\
{\mathcal{A}}_{\|}^{R}  &= N_F\, \frac{\lambda^{1/2}(m_B^2,\mdtwo^2,q^2)}{2 m_B\mdtwo} \,(m_B + \mdtwo)A_1(q^2) \,\left( - g_A^{R}\right)  \,, \\
{\mathcal{A}}_{\perp}^{L}  &=   - N_F\,\frac{\lambda(m_B^2,\mdtwo^2,q^2)}{2 m_B\mdtwo} \, \frac{ V(q^2)}{(m_B + \mdtwo)}\,\left(1 + g_V^{L}\right) \,, \\
{\mathcal{A}}_{\perp}^{R}  &=   - N_F\,\frac{\lambda(m_B^2,\mdtwo^2,q^2)}{2 m_B\mdtwo} \, \frac{ V(q^2)}{(m_B + \mdtwo)}\,g_V^{R} \,, \\
{\mathcal{A}}_{t}^L  &= N_F\, \frac{\lambda(m_B^2,\mdtwo^2,q^2)}{\sqrt6 m_B\mdtwo}\,\frac{ A_0(q^2) }{\sqrt{q^2}}\, \left(1- g_A^L\right) \,, \\
{\mathcal{A}}_{t}^R  &=  N_F\,\frac{\lambda(m_B^2,\mdtwo^2,q^2)}{\sqrt6 m_B\mdtwo}\,\frac{ A_0(q^2) }{\sqrt{q^2}}\, \left(- g_A^R\right) \,,
\end{align}
The (pseudo)scalar and tensor amplitudes can be defined as
\begin{align}
\label{tran_ampT}
{\cal{A}}_{P}^{L,R}  &=  N_F\,\frac{\lambda(m_B^2,\mdtwo^2,q^2)}{\sqrt6 m_B\mdtwo} \frac{ A_0(q^2) }{ (m_b(\mu) + m_c(\mu))}\, g_P^{L,R} \,, \\
\mathcal{A}_{T\,0}^{L,R}  &= N_F\,\frac{\lambda^{1/2}(m_B^2,\mdtwo^2,q^2)}{\sqrt6 m_B \mdtwo} \frac{1}{ \mdtwo } \Big[( m_B^2 + 3\mdtwo^2 - q^2)  T_2(q^2) \nn \\
& \hspace*{5cm} - \frac{\lambda(m_B^2,\mdtwo^2,q^2)}{m_B^2 - \mdtwo^2}  T_3(q^2) \Big]\,C^T_{LL,RR} \,, \\
\mathcal{A}_{T\,\|}^{L,R}  &= N_F\,\frac{\lambda^{1/2}(m_B^2,\mdtwo^2,q^2)}{ m_B\mdtwo} \,\frac{m_B^2 - \mdtwo^2}{\sqrt{q^2}}\, T_2(q^2)\, C^T_{LL,RR}  \,, \\
\mathcal{A}_{T\,\perp}^{L,R} &=    N_F\,\frac{\lambda(m_B^2,\mdtwo^2,q^2)}{ m_B\mdtwo}\, \frac{1}{{\sqrt{q^2}}}\, T_1(q^2) \,C^T_{LL,RR} \,.
\end{align}
The amplitudes ${\mathcal{A}}_{t}^{L,R}$ and ${\cal{A}}_{P}^{L,R}$ arise 
in a combination and hence we define
\bea
\label{tp_comb}
{\cal{A}}_{tP}^{L,R} &=& \Big({\cal{A}}_t^{L,R} + \frac{\sqrt{q^2}}{m_l} {\cal{A}}_P^{L,R} \Big)\,.
\eea

The $CP$-conjugate mode $B\to \bar{\dtwo}(\to D\pi)\ell^+ \nu$ can be described by replacing the angular coefficients $I_i$ in Eq.~\eqref{eq:dGamma} with $\bar{I}_i$ which differs by the sign flip in weak phase. Due to the change in the definition of relative angles $\theta_l\to  \theta_l - \pi$ and $\phi \to -\phi$ in the amplitudes the differential distribution $\dsp\frac{d^4\bar{\Gamma}}{dq^2d\cos\theta_D d\cos\theta_l d\phi}$ for the conjugate mode can be written with the following substitutions. 
\bea
I_{1,2,3,4,7} \to \bar{I}_{1,2,3,4,7},~~~I_{5,6,8,9} \to -\bar{I}_{5,6,8,9}.
\eea
The richness of the angular distribution is such that by performing a fit to data, each of these angular coefficients for both the mode and its conjugate mode can be extracted at experiments. This however requires more statistics and next we define several observables which individually can be accessed directly from data without going into the full fit procedure.

The differential distribution w.r.t $q^2$ can be obtained by integrating all three angles $\cos \thl,~\cos \thD$ and $\phi$ as
\bea
\label{eq:gammaF}
{d\Gamma \over dq^2} \equiv \Gamma_f= {1\over4} \left( 3I_1^c + 6I_1^s - I_2^c - 2I_2^s \right) \,.
\eea
and in absence of any direct $CP$ violation, $d{\bar\Gamma} / dq^2 \equiv \bar{\Gamma}_f = \Gamma_f$.

Integrating two angles at a time in Eq.~\eqref{eq:dGamma} generates the uniangular differential distributions. The distribution in $\cos\thD$ looks like
\begin{align}
\label{eq:dqdthD}
{d^2\Gamma \over dq^2 d\cos\thD} &= {5\over32} \big[  \left( 3I_1^c - I_2^c \right) + 6\left( -3I_1^c + I_2^c + 6 I_1^s-2 I_2^s \right) \cos^2\thD  \nn \\
&~~~~+  3\left( 9I_1^c - 3 I_2^c -12 I_1^s+4 I_2^s \right) \cos^4\thD\big] \nn \\
& = {5\over8} \Gamma_f \Big[ F_L + 6(F_T-F_L)\cos^2\thD + 3 (3 F_L-2 F_T) \cos^4\thD \Big]\,.
\end{align}
Here $F_{L,T}$ are the longitudinal and transverse polarization fractions for the $\dtwo$ meson defined as
\begin{align}
F_L= \frac{3 I_1^c -I_2^c}{ 3I_1^c + 6I_1^s - I_2^c - 2I_2^s}\,,~~F_T= \frac{2(3 I_1^s -I_2^s)}{ 3I_1^c + 6I_1^s - I_2^c - 2I_2^s}\,,
\end{align}
respectively, which satisfies $F_L+F_T=1$.
Similarly, for the $CP$-conjugate mode, one can have $\bar{F}_{L,T}$ which are equal to $F_{L,T}$, respectively, when $CP$ violation is absent. 

Now the distribution in the angle $\phi$ has a much simpler form, given as
\begin{align}
\label{eq:dqdPhi}
{d^2\Gamma \over dq^2 d\phi}&= {1\over8\pi} \big[ \left( 3I_1^c + 6I_1^s - I_2^c - 2I_2^s \right) +4 I_3 \cos2\phi+ 4 I_9 \sin2\phi \big]\, \nn \\
&= {1\over 2\pi}  \, \big[ \Gamma_f + I_3 \cos2\phi+ I_9 \sin2\phi  \big]\,,
\end{align}
where one can easily extract out the coefficients of $\cos2\phi$ and $\sin 2\phi$ terms from data. By considering the similar distribution for the $CP$-conjugate mode, we define two $CP$-averaged asymmetry $A_3$ and $A_9$ as
\bea
A_3 = \frac{I_3 + \bar{I}_3}{\Gamma_f + \bar{\Gamma}_f},~~A_9 = \frac{I_9 + \bar{I}_9}{\Gamma_f + \bar{\Gamma}_f}\,.
\eea

Integrating $\phi$ and $\cos\thD$ in Eq.~\eqref{eq:dGamma}, we get  
	\begin{align}
	{d^2\Gamma \over dq^2 d\cos\thl} &= {3\over8} \big[ \left( I_1^c + 2I_1^s - I_2^c - 2I_2^s \right) + \left( I_{6c} + 2I_{6s} \right) \cos\thl + \left( 2I_2^c + 4I_2^s \right) \cos^2\thl\big] \,.
	\end{align}

The well-known $CP$-averaged forward-backward asymmetry
$\AFB$ is defined conventionally as,
\begin{equation}
\label{eq:AFB}
\AFB=\dsp\frac{ \dsp\int_0^{2\pi}d\phi\int_{-1}^1d\cos\thD  \Big[\dsp \int_{-1}^0-\dsp\int_0^1\Big]\dsp
	d\cos\thl \frac{d^2 
		(\Gamma-\bar{\Gamma})}{d q^2d\cos\thl d\cos\thD d\phi}}{\dsp\int_0^{2\pi}d\phi\int_{-1}^1d\cos\thD \int_{-1}^1 d\cos\thl\,
	\frac{d^4(\Gamma+\bar{\Gamma})}{dq^2d\cos\thl d\cos\thD d\phi}}~,
\end{equation}

Contributions from $I_4$ and $I_5$  in Eqs.~\eqref{eq:I4} and \eqref{eq:I5} are extracted by the two angular asymmetries,
\begin{align}
  \label{eq:A4}
A_{4}=&\frac{\Big[\dsp\int_{-\pi/2}^{\pi/2}-\dsp\int_{\pi/2}^{3\pi/2} 
	\Big]d\phi
	\Big[\dsp\int_0^1 -\dsp\int_{-1}^0 \Big] d\cos\thD 
	\Big[\int_{-1}^0 - \int_0^1 \Big] d\cos\thl \dsp
	\frac{d^4(\Gamma+\bar{\Gamma})}{dq^2d\cos\thl d\cos\thD d\phi}} 
{\dsp\int_0^{2\pi}d\phi\int_{-1}^1d\cos\thD \int_{-1}^1d\cos\thl\,
	\frac{d^4(\Gamma+\bar{\Gamma})}{dq^2d\cos\thl d\cos\thD d\phi}}~,\\
\label{eq:A5}
A_{5}=&\frac{\Big[\dsp\int_{-\pi/2}^{\pi/2} -\dsp\int_{\pi/2}^{3\pi/2} 
	\Big]d\phi
	\Big[\int_0^1 -\int_{-1}^0 \Big] d\cos\thD
	\dsp\int_{-1}^1 d\cos\thl~
	\dsp\frac{d^4(\Gamma-\bar{\Gamma})}{dq^2d\cos\thl d\cos\thD d\phi} }
{\dsp\int_0^{2\pi}d\phi\int_{-1}^1d\cos\thD \int_{-1}^1d\cos\thl\,
	\frac{d^4(\Gamma+\bar{\Gamma})}{dq^2d\cos\thl d\cos\thD d\phi}}~.
\end{align}

We further define two observables $A_7$ and $A_8$ which are vanishing in the SM limit or in  other words, in real amplitude limit. These asymmetries are nonzero only
if NP introduces a complex contribution to the amplitude. A similar statement holds true for the asymmetry $A_9$ as well.
\begin{align}
\label{eq:A7} 
A_{7}=&\frac{\Big[\dsp\int_{0}^{\pi}-\int_{\pi}^{2\pi} \Big]d\phi 
	\Big[\int_0^1 -\int_{-1}^0 \Big]d\cos\thD 
	\dsp\int_{-1}^1 d\cos\thl~
	\dsp\frac{d^4(\Gamma+\bar{\Gamma})}{dq^2d\cos\thl d\cos\thD d\phi} }
{\dsp\int_0^{2\pi}d\phi\int_{-1}^1d\cos\thD \int_{-1}^1d\cos\thl\,
	\frac{d^4(\Gamma+\bar{\Gamma})}{dq^2d\cos\thl d\cos\thD d\phi}}~,\\
\label{eq:A8}
A_{8}=&\frac{\Big[\dsp\int_{0}^{\pi} -\dsp\int_{\pi}^{2\pi}\Big] d\phi 
	\Big[\dsp\int_0^1 -\dsp\int_{-1}^0 \Big]d\cos\thD  
	\Big[\int_0^1  -\int_{-1}^0 \Big] d\cos\thl \dsp
	\frac{d^4(\Gamma-\bar{\Gamma})}{dq^2d\cos\thl d\cos\thD d\phi}}
{\dsp\int_0^{2\pi}d\phi\int_{-1}^1d\cos\thD \int_{-1}^1d\cos\thl\,
	\frac{d^4(\Gamma+\bar{\Gamma})}{dq^2d\cos\thl d\cos\thD d\phi}}.
\end{align}

\section{Model independent phenomenology with New Physics}
\label{sec:pheno}

In this section, we perform a phenomenological study for the observables defined in the preceding section. The form factors for $\bar{B}\to \dtwo$ have been calculated in literature~\cite{Azizi:2013aua,Aliev:2019ojc}. In Ref.~\cite{Azizi:2013aua} a subset of form factors are estimated within a three-point QCD sum-rule approach and a recent analysis~\cite{Aliev:2019ojc} extends the previous work by providing
results for the full set of $\bar{B}\to \dtwo$ transition form factors (including the tensor form factors),  up to twist-four accuracy of $B$-meson LCSR as well as incorporating the finite virtual quark mass effects. We adopt the formalism developed in Ref.~\cite{Aliev:2019ojc} where the extrapolation of form factors from the calculated LCSR input points ($q^2 \lesssim 0\,  \gev$) to larger $q^2$ values is performed by a simple pole form with $z$-expansion 
\begin{equation}
\label{eq:fitform}
F^{B\to \dtwo}(q^2) \equiv \frac{1}{1 - q^2 / m_{R,F}^2}\, \sum_{n=0}^{1} \alpha_n^{F} \left[z(q^2) - z(0)\right]^n\,,
\end{equation}
where  $z(s) \equiv \dsp\frac{\sqrt{t_+ - s} - \sqrt{t_+ - t_0}}{\sqrt{t_+ - s} + \sqrt{t_+ - t_0}}$, $t_\pm = (m_B \pm m_{\dtwo})^2$ and $t_0 \equiv t_+ \left(1 - \sqrt{1 - t_- / t_+}\right)$.
The values for the fit parameters $\alpha_{0,1}^{F}$ are extracted in Ref.~\cite{Aliev:2019ojc} and the masses of resonances associated
with the quantum numbers of the respective form factor $F$ are taken from Ref.~\cite{Straub:2015ica}.

Only with the information about the form factors, one can easily predict the observables (discussed in the previous section) in the SM as all the Wilson coefficients $C^X_{MN}$ are vanishing in this limit. The dilepton invariant mass square varies from $m_l^2\le q^2 \le (m_B-\mdtwo)^2$. We obtain the bin-averaged value for observable $\mathcal{O}_{\rm obs}(q^2)$ defined as
\bea
\langle \mathcal{O_{\rm obs}} \rangle = \frac{\dsp\int_{q_{\rm min}^2}^{q_{\rm max}^2} dq^2\, \mathcal{O_{\rm obs}}(q^2) \,\frac{d\Gamma}{dq^2} }{\dsp \int_{q_{\rm min}^2}^{q_{\rm max}^2} dq^2 \dsp\frac{d\Gamma}{dq^2}}\,,
\eea
in different $q^2$ bins and the SM predictions for each observables with $\pm1\sigma$ uncertainties in the form factors and quark masses are shown in Table~\ref{tab:bin} for both the muon and tau mode. We divide the kinematically allowed region into eight(five) $q^2$ bins, each with $\sim1\,\gev^2$ range, for $\bar{B} \to \dtwo\, \mu(\tau) \bar{\nu}$ modes, respectively.
\begin{table}[!thb]
	\centering
		\begin{tabular}{c|c c c c c c}
			$q^2$ bin in $\gev^2$ & $\langle F_L\rangle$& $\langle F_T\rangle$ & $\langle \AFB \rangle$ & $\langle A_3\rangle$ & $\langle A_4\rangle$ & $\langle A_5\rangle$ \\[1.ex]
			\hline
			\noalign{\vskip1pt}
			\parbox[c]{4mm}{\multirow{10}{*}{\rotatebox[origin=c]{90}{$\bar{B} \to \dtwo \mu \bar{\nu}$}}} 
			$[0.1,\,1]$& $0.900 (35)$ & $0.100 (35)$  & $0.022 (25)$ & $-0.032 (10)$ & $0.033 (4)$ & $0.038 (13)$ \\[1.ex]
			~~$[1,\,2]$& $0.765 (63)$ & $0.235 (63)$  & $0.120 (52)$ & $-0.080 (19)$ & $0.050 (4)$ & $0.052 (16)$ \\[1.ex]
			~~$[2,\,3]$& $0.664 (66)$ & $0.336 (66)$  & $0.170 (61)$ & $-0.121 (22)$ & $0.056 (2)$ & $0.054 (15)$ \\[1.ex]
			~~$[3,\,4]$& $0.589 (57)$ & $0.411 (57)$  & $0.197 (62)$ & $-0.157 (21)$ & $0.060 (2)$ & $0.052 (13)$ \\[1.ex]
			~~$[4,\,5]$ & $0.531 (44)$  & $0.469 (44)$ & $0.204 (59)$ & $-0.189 (18)$ & $0.061 (1)$ & $0.047 (12)$ \\[1.ex]
			~~$[5,\,6]$ & $0.485 (31)$  & $0.515 (31)$ & $0.196 (53)$ & $-0.221 (15)$ & $0.062 (1)$ & $0.040 (10)$ \\[1.ex]
			~~$[6,\,7]$ & $0.448 (18)$  & $0.552 (18)$  & $0.169 (44)$& $-0.251 (10)$ & $0.063 (1)$ & $0.032 (8)$ \\[1.ex]
			~~$[7,\,7.9]$ & $0.420 (6)$  & $0.580 (6)$   & $0.120 (27)$& $-0.278 (4)$ & $0.063 (0)$ & $0.021 (4)$ \\[1.5ex] \hline 
			\noalign{\vskip2pt}
			\parbox[c]{4mm}{\multirow{6}{*}{\rotatebox[origin=c]{90}{$\bar{B} \to \dtwo \tau \bar{\nu}$}}} 
			$[3.1,\,4]$& $0.662 (32)$ & $0.338 (32)$  & $-0.179 (38)$ & $-0.015 (1)$ & $0.006 (0)$ & $0.076 (11)$ \\[1.ex]
			~~$[4,\,5]$ & $0.600 (29)$  & $0.401(29)$  & $-0.099 (36)$ & $-0.038 (4)$ & $0.012 (1)$ & $0.071 (11)$ \\[1.ex]
			~~$[5,\,6]$ & $0.533 (22)$  & $0.467 (22)$  & $-0.028 (33)$ & $-0.067 (5)$ & $0.019 (1)$ & $0.062 (11)$ \\[1.ex]
			~~$[6,\,7]$ & $0.475 (14)$  & $0.525 (14)$  & $0.017 (28)$& $-0.099 (5)$ & $0.025 (1)$ & $0.049 (9)$ \\[1.ex]
			~~$[7,\,7.9]$ & $0.431 (5)$  & $0.569 (5)$   & $0.031 (18)$& $-0.128 (2)$ & $0.029 (0)$ & $0.033 (5)$ \\[1.ex]
			\hline
			\hline
		\end{tabular}
		\caption{The SM predictions for the bin-averaged values of the observables with $\pm 1\sigma$ uncertainty for $\bar{B} \to \dtwo \mu \bar{\nu}$ and $\bar{B} \to \dtwo \tau \bar{\nu}$ decays. 	}
		\label{tab:bin}
\end{table}

\begin{figure*}[!h]
	\begin{center}
		\includegraphics[width=0.45\linewidth]{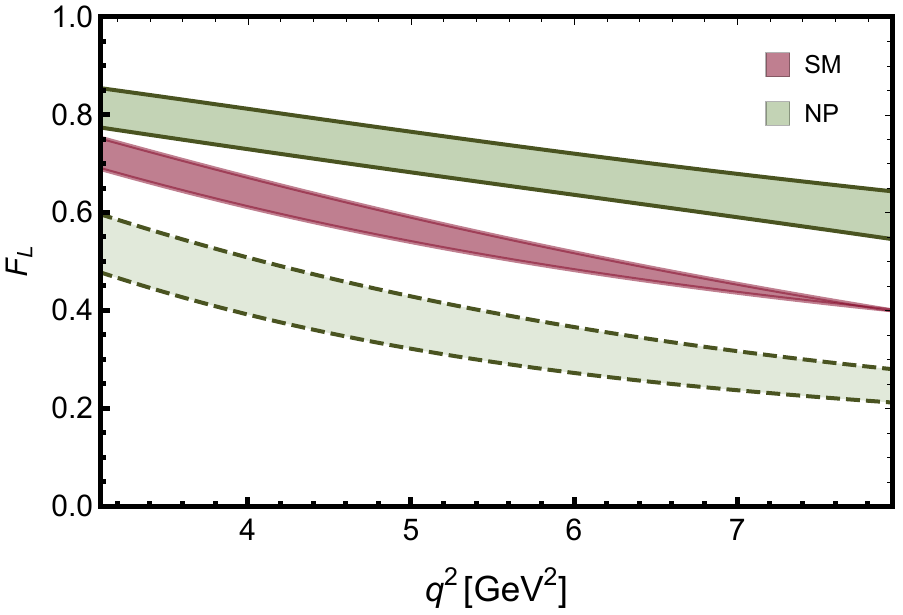} \hskip 10pt
		\includegraphics[width=0.45\linewidth]{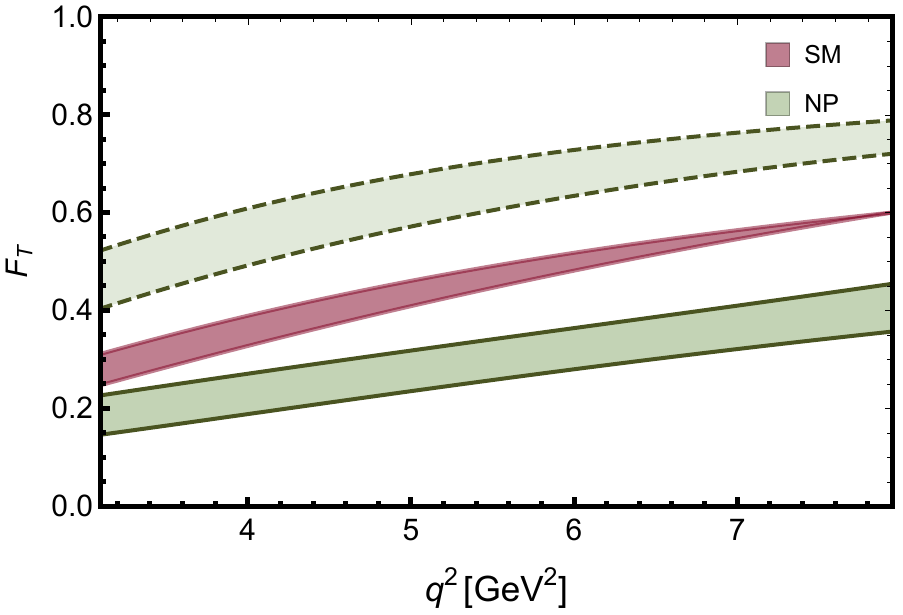}
		\includegraphics[width=0.46\linewidth]{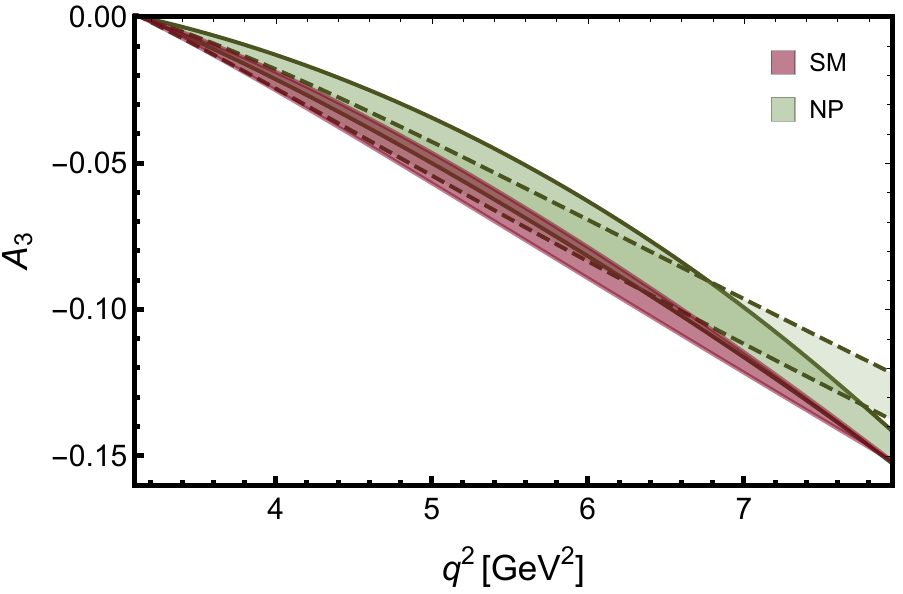} \hskip 10pt
		\includegraphics[width=0.45\linewidth]{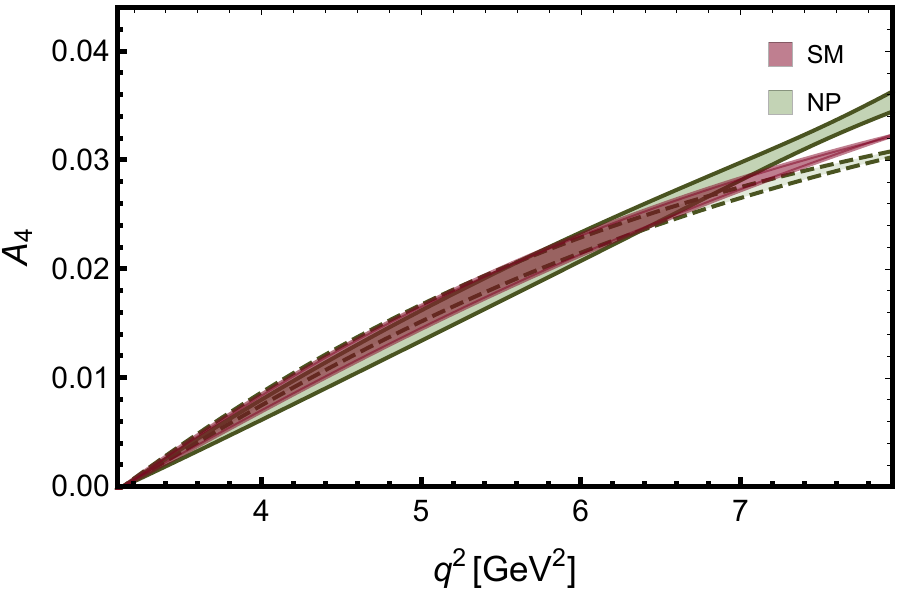}
		\includegraphics[width=0.46\linewidth]{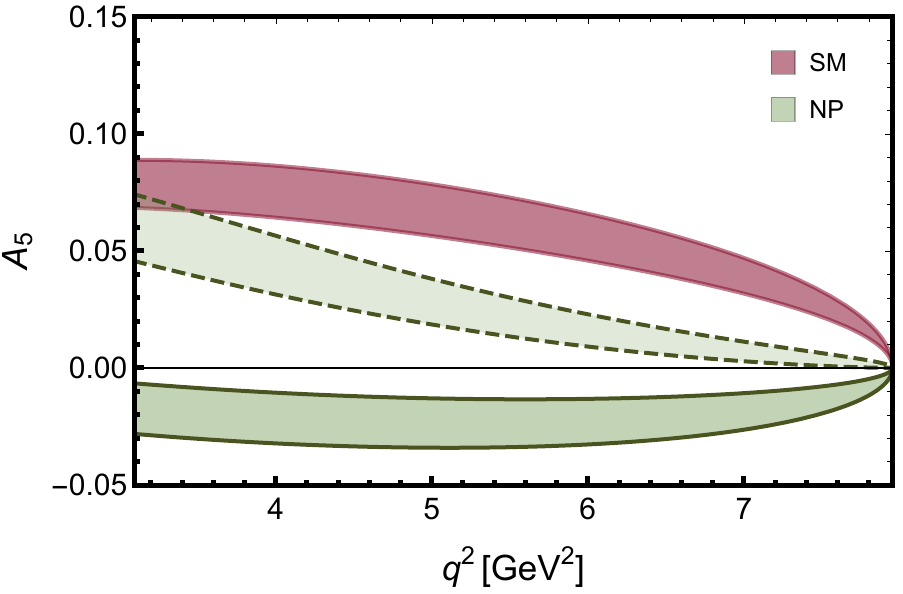} \hskip 10pt
		\includegraphics[width=0.45\linewidth]{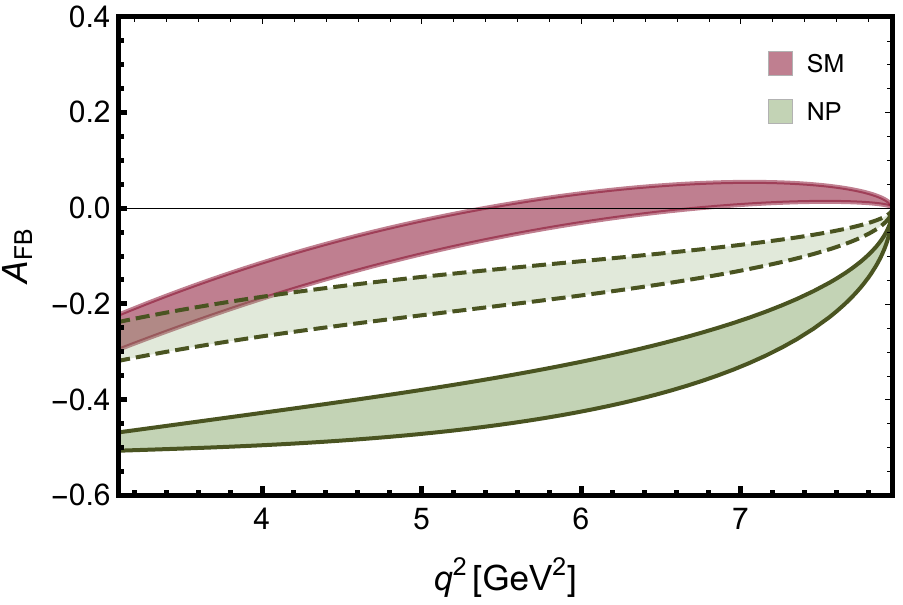}
		\caption{The variation of the observables for $\bar{B} \to \dtwo(\to D \pi) \tau \bar{\nu}$ decay in the allowed $q^2$ range where the red bands correspond to the SM case. The solid and dashed green bands denote the NP contributions given in Eq.~\eqref{eq:WCfit} and Eq.~\eqref{eq:WCfitM}, respectively, which correspond to the two minima of the global fit~\cite{Murgui:2019czp} to $b\to c \tau \bar{\nu}$ data.}\label{fig:Obs}
	\end{center}
\end{figure*}

In Fig.~\ref{fig:Obs} we show the variation of these observables, in the entire $q^2$ range, in red bands for the SM prediction of the $\bar{B} \to \dtwo \tau \bar{\nu}$ mode. Note that the forward-backward asymmetry $\AFB$ has a zero crossing in the $q^2$ axis. It implies there exists a relation between the form factors involved in the matrix elements in the SM limit at $\AFB(q_0^2)=0$ which is given by
\bea
3 A_1(q_0^2) V(q_0^2) = \frac{m_\tau^2}{q_0^4} \frac{A_0(q_0^2)}{\mdtwo} \left[ A_1(q_0^2) (m_B+\mdtwo) (m_B^2 -\mdtwo^2 -q_0^2) - \frac{\lambda (m_B^2,\mdtwo^2,q_0^2)}{m_B + \mdtwo}  A_2(q_0^2) \right].
\eea
With the estimates of the central values of the form factors, we find $q_0^2= 6.0 \,\gev^2$. As it can be seen from Eq.~\eqref{eq:AFB}  that the above relation originates from the cancellation between $I_6^c$ and $2 I_6^s$ terms, such cancellation is absent in $\bar{B} \to \dtwo \mu \bar{\nu}$ mode due to the low mass of the muon compared to tau. Interestingly in the presence of the NP operators the relation will be modified with the Wilson coefficients $C^X_{MN}$ and the tensor form factors $T_i(q^2)$ and will have a bit lengthy form. However the relation becomes simpler with NP contribution (with real Wilson coefficients) to only (axial)vector operators which can be written as
\begin{align}
\frac{(1- g_A^L)(1+ g_V^L) + g_A^R g_V^R}{(1- g_A^L)^2 + {g_A^R}^2} =&\, \frac{m_l^2}{q_0^4}\frac{A_0(q_0^2)}{3V(q_0^2)A_1(q_0^2)}\,\frac{1}{\mdtwo} \times \nn \\  &\left[ A_1(q_0^2) (m_B+\mdtwo) (m_B^2 -\mdtwo^2 -q_0^2) - \frac{\lambda (m_B^2,\mdtwo^2,q_0^2)}{m_B + \mdtwo}  A_2(q_0^2) \right].
\end{align}
Hence we infer that if the zero crossing point of $\AFB$ is measured in the future, it will provide important information about the NP Wilson coefficients.

Next we test the sensitivity of the observables in the presence of NP contributions. The intriguing discrepancies in the $b\to c \tau \bar{\nu}$ transition have predicted nonzero value(s) for one and/or several NP Wilson coefficients. We follow one of the most recent model independent analyses where global fit to the general set of Wilson coefficients of an effective low-energy Hamiltonian (with only left-handed neutrinos) is performed~\cite{Murgui:2019czp}. As a benchmark scenario, we consider the ``Min 4" from Table 6 of \cite{Murgui:2019czp}, where $\chi^2$ minimization to $R(D^{(*)})$, $D^*$ longitudinal polarization fraction $F_L^{D^*}$, and binned $q^2$ distributions for $B \to D^{(*)} \tau \bar{\nu}$ data is presented for five NP coefficients. The bound from ${\rm BR}(B_c \to \tau \bar{\nu})\le 10\%$ has also been imposed in the fit. We consider the central values of the fitted Wilson coefficients and using Eq.~\eqref{eq:WC-nom} the combination of coefficients entering in the tranversity amplitudes are found to be
\bea
\label{eq:WCfit}
g_V^L=0.98,~g_A^L=2.8,~g_P^L=0.90,~C_{LL}^T=-0.22\,.
\eea
The observables with the above-mentioned NP coefficients are shown in solid green bands in Fig.~\ref{fig:Obs} where all other Wilson coefficients are assumed to be vanishing.

It was mentioned in Ref.~\cite{Murgui:2019czp} that a degeneracy between the set of Wilson coefficients and a sign-flipped minimum exists corresponding to the same minimized $\chi^2$ value. For such case, the values in Eq.~\eqref{eq:WCfit} alter to
\bea
\label{eq:WCfitM}
g_V^L=0.8,~g_A^L=2.98,~g_P^L=-0.90,~C_{LL}^T=0.22\,.
\eea
In Fig.~\ref{fig:Obs} we highlight the variation of observables corresponding to this mirror minima in dashed green bands. It can be seen that the difference in two NP scenarios (which are indistinguishable in terms of minimization) is quite prominent for the helicity fractions $F_L$, $F_T$ and asymmetries $A_5$ and $\AFB$. However, $A_3$ and $A_4$ show alteration only near the kinematic endpoint.

It should be noted that the angular observables $A_7,~A_8$ and $A_9$ depend on the imaginary part of the transversity amplitudes and therefore are vanishing in the SM. Only NP contributions with complex Wilson coefficients can make them finite and hence measurements of nonzero values of these  observables will be a clean signal of NP contributing to this decay mode.

\section{Summary and discussion}
\label{sec:summary}

In this paper, we have explored the semileptonic decay $\bar{B}\to \dtwo (\to D \pi) \ell \bar{\nu}$, where $\dtwo$ is a tensor meson with a mass and decay width of $2460$\,MeV and $\sim 48$\,MeV, respectively. We start with the most general beyond the SM effective Hamiltonian in dimension-six operator basis which comprises (axial)vector, (pseudo)scalar and tensor operators for quark as well as lepton currents. We also include the right-handed neutrinos in the effective Hamiltonian. 

The further decay of $\dtwo \to D \pi$ states allows us to derive the full four-fold angular distribution. The entire decay can be expressed in terms of 16 transversity amplitudes while in the SM limit the number of transversity amplitudes simply reduces to four. We find the NP contribution to (axial)vector operators can be incorporated in the SM amplitudes modified with the NP Wilson coefficients; however, the (pseudo)scalar and tensor operators induce four new transversity amplitudes for each chirality of the lepton current.

A multitude of $CP$-averaged observables is constructed from the full differential distribution. These are the helicity fractions of the $\dtwo$ meson $F_L,~F_T$, the forward-backward asymmetry $\AFB$ and six angular asymmetries $A_i,~i\in\{3,4,5,7,8,9\}$ which can be extracted at experiments. Among these, three asymmetries $A_{7,8,9}$ depend only on the complex part of the decay amplitude and hence are vanishing in the SM as well as real NP Wilson coefficient limit.

Next we predict the bin-averaged values of the observables for several $q^2$ bins for $\bar{B}\to \dtwo (\to D \pi) \mu \bar{\nu}$ and $\bar{B}\to \dtwo (\to D \pi) \tau \bar{\nu}$ channels in the SM. We also illustrate the behavior of the observables in the entire kinematical allowed range in the presence of NP contributions which can explain the intriguing discrepancies observed in $b\to c \tau \bar{\nu}$ transitions. By using the results from the latest global fit to all the relevant observables, we find in some observables e.g., $F_L,~F_T,~A_5$ and $\AFB$, the effects arising from the NP contributions are quite prominent. We show the zero crossing of the forward-backward asymmetry $\AFB$ can also provide important information about the NP coefficients.

The results derived in this work are not only restricted to the particular decay but can also be applied to other channels with tensor meson in the final state like $\bar{B}_s \to D_{2s}^* (\to D K) \ell \bar{\nu}$ where $D_{2s}^*(2573)$ is a $2^+$ state. The experimental sensitivity to perform a fit to the full four-fold distribution for extracting each angular coefficient for these modes is subject to the statistics; however, we hope the observables constructed in this work can be tested at the LHC and Belle-II experiments in the near future.  

\subsection*{Acknowledgment}
We thank Antonio Pich for useful discussions.
This work has been supported by the Alexander von Humboldt Foundation through a postdoctoral research fellowship.

\appendix
\section{Decay kinematics}
\label{app:kin}

In this section we quote the expressions for the kinematics which are used in squaring the matrix element of the decay $\bar{B}(p_B)\to \dtwo(p_{\dtwo})\,\ell^-(q_2)\,\bar{\nu}(q_1)$, with $\dtwo(p_{\dtwo})\to
D(p_D)\pi(p_\pi)$ on the mass shell.

With the notation introduced in Eq.~\eqref{eq:notation} and defining $Q^\mu=q_1^\mu-q_2^\mu$ we list the following expressions.
\begin{align}
p_{\dtwo}.q&= \frac{m_B^2-q^2-p_{\dtwo}^2}{2}\,, \\
p_{\dtwo}.Q&= \alpha_l X \cos\thl-\frac{m^2}{q^2}p_{\dtwo}.q\, ,\\
q.P &=\xi  (p_{\dtwo}.q)- \beta  X \cos \thD, \\
q.Q &= -m^2, \\
Q^2 &= 2 m^2-q^2, \\
P^2 &= k^2 \left(\xi^2-\beta^2\right),  \\
p_{\dtwo}.P &= \xi \mdtwo^2\,,  \\
P.Q &=   \xi  (p_{\dtwo}.Q) -\alpha_l \beta (p_{\dtwo}.q) \cos \thl  \cos \thD  + \frac{m^2}{q^2}\beta X \cos \thD \nn \\
&-  \alpha_l\beta \sqrt{q^2}\, \mdtwo \sin \thl  \sin \thD \cos \phi\,, \\
\epsilon_{\mu\nu\alpha\beta} p_{\dtwo}^\mu P^\nu q^\alpha Q^\beta &= -  \alpha_l\beta \sqrt{q^2}\, \mdtwo  X \sin\thl \sin \thD \sin \phi\,.
\end{align}
where $\alpha_l= \left( 1- \dsp\frac{m_l^2}{q^2}\right),~X= \dsp\frac{\lambda^{1/2}(m_B^2,\mdtwo^2,q^2)}{2} $ and $\beta=\dsp\frac{\lambda^{1/2}(\mdtwo^2,m_D^2,m_\pi^2)}{\mdtwo^2} $.

\end{document}